\documentclass[aps,prd,onecolumn,groupedaddress,showpacs,nofootinbib,amssymb]{revtex4}
\usepackage[dvips]{graphicx}
\usepackage{amssymb}
\usepackage{amsmath}
\usepackage{graphicx,,color}
\usepackage{amsfonts}
\usepackage{bm}
\usepackage{cancel}
\usepackage{comment}

\newcommand\be{\begin{equation}}
\newcommand\ee{\end{equation}}

\allowdisplaybreaks[4]

\begin{document}

\title{Rectifying an Inconsistency in $F(R)$ Gravity Inflation}
\author{V.K. Oikonomou,$^{1,2,3}$\,\thanks{v.k.oikonomou1979@gmail.com}}
\affiliation{$^{1)}$Department of Physics, Aristotle University of Thessaloniki, Thessaloniki 54124, Greece\\
$^{2)}$ Laboratory for Theoretical Cosmology, Tomsk State
University of Control Systems
and Radioelectronics, 634050 Tomsk, Russia (TUSUR)\\
$^{3)}$ Tomsk State Pedagogical University, 634061 Tomsk, Russia }

\tolerance=5000

\begin{abstract}
In this letter we shall provide a formal derivation of the
inflationary slow-roll indices for $F(R)$ gravity from the power
spectrum without the condition $\dot{\epsilon}_1=0$, frequently
used in the literature, where $\epsilon_1=-\frac{\dot{H}}{H^2}$ is
the first slow-roll index. We shall employ Karamata's theorem for
regularly varying functions, and as we show, we shall derive the
same expressions for the scalar and tensor spectral index of
$F(R)$ gravity, also appearing in the current literature, without
the misleading and rather strong condition $\dot{\epsilon}_1=0$.
The only conditions that are needed are the slow-roll assumption,
and the smallness of the slow-roll indices and of their
derivatives during inflation.
\end{abstract}


\maketitle

The primordial era of our Universe remains still a mystery that
needs to be resolved by modern theoretical cosmology. During that
primordial era, the Universe emerged from a hypothetical quantum
era, if this ever existed, and spacetime was four dimensional as
we know it today, with most particle species already existing at
that stage. Then it is hypothesized that the Universe went through
a rapid acceleration era, the so-called inflationary era. This
inflationary process solved theoretically the problems of the
standard Big Bang scenario, such as the flatness and horizon
problems, but even up to date we have not confirmed that this era
indeed occurred in the way we theorize that it must have occurred.
What we have at hands, via the Cosmic Microwave Background
radiation observational data, is that the power spectrum of the
primordial scalar curvature perturbations is nearly flat, but
still this feature does not confirm the fact that the inflationary
era indeed occurred. A direct confirmation of the inflationary era
will be the verification of the $B$-modes in the CMB
\cite{Kamionkowski:2015yta}, but this might be a far future
observation.

For the description of the inflationary era, there are many
scenarios that can harbor such a physical process, with most of
them relying on scalar fields
\cite{inflation1,inflation2,inflation3,inflation4}. However, in
nature we have observed directly only one scalar field, the Higgs
boson, which basically gives mass to all the known massive
particles via the spontaneous symmetry breaking mechanism. Apart
from the Higgs boson, no other scalar particle has ever been
observed. Apart from the scalar field description of the
inflationary scenario, the modified gravity description stands out
as a phenomenologically appealing alternative description
\cite{reviews1,reviews2,reviews3,reviews4}, with the most
prominent modified gravity theory being the $F(R)$ gravity theory.
Especially interesting is the fact that in $F(R)$ gravity, the
unification of dark energy with inflation is possible, see for
example \cite{Nojiri:2006ri,Nojiri:2003ft}. In the context of
$F(R)$ gravity, the inflationary era is realized by higher order
scalar curvature terms. Actually, $F(R)$ gravity modifies the
right hand side of the Einstein equations, introducing an
effective energy momentum tensor entirely generated by geometric
terms. Among many $F(R)$ gravity models that may yield
phenomenologically viable results \cite{reviews1} compatible even
with the latest Planck data \cite{Akrami:2018odb}, the $R^2$ model
\cite{Starobinsky:1980te} model is the most well-known, see also
\cite{Bezrukov:2007ep} for the non-minimally coupled scalar theory
analogue.

Nowadays, the precision cosmology scientific era compels every
theoretical model to be compatible with observational data. To
this end, for the $F(R)$ gravity description of inflation, there
exist solid steps for the calculation of the spectral index of the
primordial scalar curvature and tensor perturbations and of the
tensor-to-scalar ratio, and these can be expressed in terms of the
slow-roll indices. However, in the literature there exists an
inconsistency in the derivation of the spectral index in terms of
the slow-roll indices, and the inconsistency is spotted directly
when one tries to extract the observational indices from the power
spectrum. Particularly, this inconsistency is related to the
condition $\dot{\epsilon}_1=0$, where $\epsilon_1$ is the first
slow-roll index $\epsilon_1=-\frac{\dot{H}}{H^2}$, that is assumed
to hold true during some intermediate step for the derivation of
the spectral index in terms of the slow-roll indices. If this
condition is taken into account, even the $R^2$ model may rendered
non-viable. In this letter, we shall demonstrate that this
condition is in fact not necessary for the calculation, and by
using Karamata's \cite{karamata} theorem, we shall provide a more
refined version of the $F(R)$ inflationary dynamics. As we show
the resulting expressions for the scalar and tensor spectral index
are the same with those in the literature, without taking into
account the condition $\dot{\epsilon}_i=0$. Thus we provide a
formally more correct derivation of the expressions that yield the
observational indices of inflation for $F(R)$ gravity theories.

We shall consider inflation of $F(R)$ gravity theories from ground
zero, and we shall critically discuss all the conditions that are
needed in order to express the observational indices of inflation
in terms of the slow-roll indices. Particularly, the role of the
condition $\dot{\epsilon}_1\neq 0$, where $\epsilon_1$ is the
first slow-roll index $\dot{\epsilon}_1=-\frac{\dot{H}}{H^2}$ is
going to be critically discussed.

The $F(R)$ gravity theory in vacuum has the following action,
\begin{equation}\label{action1dse}
\mathcal{S}=\frac{1}{2\kappa^2}\int \mathrm{d}^4x\sqrt{-g}F(R),
\end{equation}
where $\kappa^2$ is defined to be $\kappa^2=8\pi
G=\frac{1}{M_p^2}$ and $M_p$ stands for the reduced Planck mass.
We shall employ the metric formalism, so the field equations are
obtained by varying the gravitational action with respect to the
metric tensor, so we obtain,
\begin{equation}\label{eqnmotion}
F_R(R)R_{\mu \nu}(g)-\frac{1}{2}F(R)g_{\mu
\nu}-\nabla_{\mu}\nabla_{\nu}F_R(R)+g_{\mu \nu}\square F_R(R)=0\,
,
\end{equation}
with $F_R=\frac{\mathrm{d}F}{\mathrm{d}R}$. The above equation can
be cast as follows,
\begin{align}\label{modifiedeinsteineqns}
R_{\mu \nu}-\frac{1}{2}Rg_{\mu
\nu}=\frac{\kappa^2}{F_R(R)}\Big{(}T_{\mu
\nu}+\frac{1}{\kappa^2}\Big{(}\frac{F(R)-RF_R(R)}{2}g_{\mu
\nu}+\nabla_{\mu}\nabla_{\nu}F_R(R)-g_{\mu \nu}\square
F_R(R)\Big{)}\Big{)}\, .
\end{align}
Assuming that the geometric background is a flat
Friedmann-Robertson-Walker (FRW) metric with line element,
\begin{equation}
\label{JGRG14} ds^2 = - dt^2 + a(t)^2 \sum_{i=1,2,3}
\left(dx^i\right)^2\, ,
\end{equation}
the field equations (\ref{modifiedeinsteineqns}) take the
following form,
\begin{align}
\label{JGRG15} 0 =& -\frac{F(R)}{2} + 3\left(H^2 + \dot H\right)
F_R(R) - 18 \left( 4H^2 \dot H + H \ddot H\right) F_{RR}(R)\, ,\\
\label{Cr4b} 0 =& \frac{F(R)}{2} - \left(\dot H +
3H^2\right)F_R(R) + 6 \left( 8H^2 \dot H + 4 {\dot H}^2 + 6 H
\ddot H + \dddot H\right) F_{RR}(R) + 36\left( 4H\dot H + \ddot
H\right)^2 F_{RRR} \, ,
\end{align}
where $F_{RR}=\frac{\mathrm{d}^2F}{\mathrm{d}R^2}$, and
$F_{RRR}=\frac{\mathrm{d}^3F}{\mathrm{d}R^3}$. Moreover, the
Hubble rate is $H=\dot a/a$, while the Ricci scalar for the FRW
metric is $R=12H^2 + 6\dot H$.

Now we shall concentrate on the inflationary phenomenology of
$F(R)$ gravity theories, and the slow-roll indices quantify
perfectly the dynamics of inflation. For $F(R)$ theories, these
are defined as follows \cite{Hwang:2005hb,reviews1},
\begin{equation}
\label{restofparametersfr}\epsilon_1=-\frac{\dot{H}}{H^2}, \quad
\epsilon_2=0\, ,\quad \epsilon_3= \frac{\dot{F}_R}{2HF_R}\, ,\quad
\epsilon_4=\frac{\ddot{F}_R}{H\dot{F}_R}\,
 .
\end{equation}
Also we shall assume that the slow-roll conditions hold true,
which in terms of the Hubble rate, these read,
\begin{equation}\label{slowrollconditionshubble}
\ddot{H}\ll H\dot{H},\,\,\, \frac{\dot{H}}{H^2}\ll 1\, .
\end{equation}
Actually, the second condition guarantees that the inflationary
era occurs in the first place. Also in the following we shall
assume that the slow-roll indices and their derivatives during
inflation, have small values, a quantification of the slow-roll
conditions. We need to note that these assumptions are purely
physically motivated and are not mathematically motivated
conditions. The slow-roll conditions guarantee that the
inflationary era occurs and that it lasts long enough. However, as
we show, the slow-roll conditions make the slow-roll indices
slowly varying during the inflationary era, and this feature
affects also in a mathematical way the results of this paper, in
terms of Karamata's theorem that we will use.

Let us now get to the core of this paper, and we aim to express
the scalar spectral index of the primordial curvature
perturbations $n_{\mathcal{S}}$, the tensor spectral index $n_T$,
and the tensor-to-scalar ratio $r$ in terms of the slow-roll
indices. We shall begin from ground zero, so we shall follow Ref.
\cite{Hwang:2005hb}, and we also modify the notation of Ref.
\cite{Hwang:2005hb} to comply with our notation. The parameters
$z$ and $z_t$ that enter the scalar and curvature perturbations
differential equations are defined as follows \cite{Hwang:2005hb},
\begin{equation}\label{perturbationeqnsparameters}
z=\frac{a}{(1+\epsilon_3)H}\sqrt{\frac{E}{F_R}}\, ,
\end{equation}
\begin{equation}\label{perturbationeqnsparameters1}
z_t=a\sqrt{F_R}\, ,
\end{equation}
where $a$ is the scale factor, and the parameter $E$ in the case
of vacuum $F(R)$ gravity is defined as follows,
\begin{equation}\label{parameterE}
E=\frac{3\dot{F}_R}{2\kappa^2}\, .
\end{equation}
Then, the derivatives of $z$ and $z_t$ with respect to the
conformal time $\eta$ are expressed in terms of the slow-roll
indices as follows,
\begin{align}\label{firstbigequation}
&
\frac{z''}{z}=a_c^2H_c^2(1+\epsilon_1+\epsilon_2-\epsilon_3+\epsilon_4)(2+\epsilon_2-\epsilon_3+\epsilon_4)+a^2H(\dot{\epsilon}_1+\dot{\epsilon}_2-\dot{\epsilon}_3+\dot{\epsilon}_4)\\
\notag &
-2a_c^2H_c\left(\frac{3}{2}+\epsilon_1+\epsilon_2-\epsilon_3+\epsilon_4\right)\frac{\dot{\epsilon}_3}{1+\epsilon_3}-\frac{a^2\ddot{\epsilon}_3}{1+\epsilon_3}+\frac{2a^2\dot{\epsilon}_3^2}{(1+\epsilon_3)^2}\,
,
\end{align}
\begin{equation}\label{secondbigequation}
\frac{z_t''}{z}=a_c^2H_c^2\left((1+\epsilon_3)(2-\epsilon_1+\epsilon_3)+\frac{\dot{\epsilon}_3}{H_c}
\right)\, ,
\end{equation}
where $a_c$ and $H_c$ are the scale factor and the Hubble rate at
horizon crossing respectively, and also it is important to note
that also the slow-roll indices must be evaluated at the horizon
crossing time instance.

At this point, the authors of Ref. \cite{Hwang:2005hb} make a
crucial assumption, namely that $\dot{\epsilon}_1=0$, or
alternatively put in the literature \cite{Stewart:1993bc},
$\epsilon_1=$const. However, this has a crucial impact on the
phenomenology of $F(R)$ gravities, since it crucially forbids to
have $\dot{\epsilon}_1\neq 0$. Actually, this never occurs for
most of the $F(R)$ gravities of interest. However, as we now show,
the only assumptions needed are that $\epsilon_i\ll 1$, $i=1,3,4$
and $\dot{\epsilon}_i\ll 1$, $i=1,3,4$.

The authors of Ref. \cite{Hwang:2005hb} by making the assumption
$\dot{\epsilon}_1=0$ result into the equation,
\begin{equation}\label{etarelation}
\eta=-\frac{1}{a_cH_c}\frac{1}{1-\epsilon_1}\, .
\end{equation}
We now show, using only the slow-roll assumption, that the above
relation indeed holds true, without assuming that
$\dot{\epsilon}_1=0$. The conformal time is defined as,
\begin{equation}\label{conf1}
d\eta=\frac{dt}{a}\, ,
\end{equation}
so by integrating we easily obtain,
\begin{equation}\label{conf2}
\eta_0-\eta=\int_{a_c}^{a_e}\frac{d a}{a^2 H}\, ,
\end{equation}
where $a_e$ is the scale factor at the end of the inflationary
era, and $a_c$ is assumed to be the scale factor at the horizon
crossing. We choose the initial value of the scale factor to be
exactly on the horizon crossing time instance, since it is exactly
then that the slow-roll indices must be evaluated in order to
calculate the spectral indices and the tensor-to-scalar ratio.


We shall use Karamata's theorem for regularly varying functions
\cite{karamata},
\begin{equation}\label{karamata}
\int_{t}^{\infty}t^{\gamma}f(t)=-(\gamma+1)^{-1}tf(t)\, ,
\end{equation}
for $t\to \infty$ and $\gamma<-1$, and $f(t)=t^{\gamma}L(t)$,
where $L(t)$ is a slowly varying function.  By integrating by
parts in Eq. (\ref{conf2}) we get,
\begin{equation}\label{conf4}
\eta_0-\eta=\int_{a_c}^{a_e}\frac{d a}{a^2
H}=-\frac{1}{a_eH_e}+\frac{1}{a_cH_c}+\int_{a_c}^{a_e}\frac{\epsilon_1}{a^2H}da\,
,
\end{equation}
where $H_e$ is the Hubble rate at the end of inflation, $H_c$ the
Hubble rate at horizon crossing, $\eta_0$ the conformal time at
the end of inflation, and $\eta$ the conformal time at horizon
crossing. Now, the function $f(a)=\frac{\epsilon_1(a)}{a^2 H}$ is
a regularly varying function, and the Hubble rate and the
slow-roll index $\epsilon_1$ are slowly varying functions during
the inflationary era. Actually for an exactly de Sitter evolution,
the Hubble rate is constant, and therefore for a quasi-de Sitter
evolution, the Hubble rate is also a slowly varying function and
the same applies for the slow-roll index $\epsilon_1$. Taking
$\eta_0=-\frac{1}{a_eH_e}$ and using Karamata's theorem for the
integral in the right hand side of Eq. (\ref{conf4}) we get,
\begin{equation}\label{conf5}
\eta=-\frac{1}{a_cH_c}\left(1+\epsilon_1(a_c) \right)\, ,
\end{equation}
where for the application of Karamata's theorem, the slowly
varying function is $L(a)=\frac{\epsilon_1}{H}$, so we obtained,
\begin{equation}\label{integralactual}
\int_{a_c}^{a_e}\frac{\epsilon_1}{a^2H}da\sim
\frac{\epsilon_1(a_c)}{a_cH_c}\, ,
\end{equation}
when $a_c$ takes large values. The above holds true approximately,
since $a_e\gg a_c$ and also $a_e$ is considered to occur well
after $10$ $e$-foldings, so $N=\ln \frac{a_c}{a_i}=10$. Hence
$a_i\ll a_c$ where recall $a_c$ is the scale factor at the horizon
crossing, and $a_c\ll a_e$, but still $a_c$ is quite large. What
Karamata's theorem tells in simple words, thus formalizing an
obvious assessment for the behavior of the integral, is that the
regularly varying function $f(a)=\frac{\epsilon_1(a)}{a^2 H}$ when
integrated over in the interval $[a_c,a_e]$, with $a_e\gg a_c$
(actually $a_e\to \infty$), the integration is determined solely
from the power-law term $a^{-2}$, and not from the slowly varying
function $\epsilon_1/H$ which contributes solely a constant to be
equal to the value of the slowly-varying function at the lower
limit of the integration (this is actually the definition of a
slowly-varying function). The power-law term when integrated, the
resulting expression at infinity (at $a_e$) goes to zero
approximately and the integral itself is mainly controlled by the
power-law part of the regularly varying function at the lower
limit of the integration. Particularly, the value of the integral
is evaluated approximately at the lower limit of the interval
$[a_c,\infty]$.

Coming back to the problem at hand, since $\epsilon_1\ll 1$, we
have,
\begin{equation}\label{taylorofepsilon1}
\frac{1}{1-\epsilon_1}\simeq
1+\epsilon_1+\mathcal{O}(\epsilon_1^2)\, ,
\end{equation}
therefore, Eq. (\ref{conf5}) becomes,
\begin{equation}\label{finalconformatimerelation}
\eta=-\frac{1}{a_cH_c}\frac{1}{1-\epsilon_1(a_c)}\, ,
\end{equation}
which is functionally identical to Eq. (\ref{etarelation}), but
instead of $\epsilon_i(a_c)$ from now on we simply write
$\epsilon_i$ for simplicity. We need to note that if the above
approach is not adopted, then the result of Ref.
\cite{Hwang:2005hb} merely holds true for $\epsilon_1=$const,
which clearly is an insufficient condition, which cannot describe
most of the interesting $F(R)$ gravity theories. So with our
approach we refined the argument of Ref. \cite{Hwang:2005hb}, by
clearly allowing the conditions $\epsilon_1(a_c)\neq $const and
$\dot{\epsilon}_1(a_c)\neq 0$. Let us now proceed to the
expressions of the spectral index $n_{\mathcal{S}}$ and of the
tensor-to-scalar ratio $r$ as functions of the slow-roll
parameters.

In Eq. (\ref{firstbigequation}) and (\ref{secondbigequation}), the
leading order terms at horizon crossing when the scale factor was
$a=a_c$, are $\sim H_c^2\epsilon_i(\alpha_{c})$, with $i=1,2,3,4$,
while the terms,
\begin{align}\label{subleadingorder}
& \sim \dot{\epsilon}_i(a_c)\, , \,\,\, \sim H_c^2\epsilon_i(a_c)\epsilon_j(a_c)\\
\notag & \sim \ddot{\epsilon}_i(a_c)\, , \,\,\, \sim
\dot{\epsilon}_i^2(a_c)
\end{align}
are subdominant at leading order, having assumed that the
slow-roll conditions $\epsilon_i(a_c)\ll 1$ hold true, and
additionally that $\dot{\epsilon}_i(a_c)\ll \epsilon_i(a_c)$.
Therefore, the terms $\frac{z''}{z}$ and $\frac{z_t''}{z_t}$ at
the horizon crossing time instance, in view of the simplifications
(\ref{subleadingorder}) and also in view of the relation
(\ref{finalconformatimerelation}) can be approximated as follows,
\begin{equation}\label{derivatives1}
\frac{z''}{z}=\frac{n_s}{\eta^2}=\frac{1}{\eta^2}\frac{1}{(1-\epsilon_1(a_c))^2}\left(2+2\epsilon_1(a_c)-3\epsilon_3(a_c)+3\epsilon_4(a_c)
\right)\, ,
\end{equation}
\begin{equation}\label{derivatives2}
\frac{z_t''}{z_t}=\frac{n_t}{\eta^2}=\frac{1}{\eta^2}\frac{1}{(1-\epsilon_1(a_c))^2}\left(2-\epsilon_1(a_c)+3\epsilon_3(a_c)\right)\,
,
\end{equation}
where the parameter $n_s$ (not to be confused with the spectral
index $n_{\mathcal{S}}$) is defined as follows,
\begin{equation}\label{nsparameter}
n_s=\frac{1}{(1-\epsilon_1(a_c))^2}\left(2+2\epsilon_1(a_c)-3\epsilon_3(a_c)+3\epsilon_4(a_c)
\right)\, ,
\end{equation}
and the parameter $n_t$ is defined as follows,
\begin{equation}\label{ntparameter}
n_t=\frac{1}{(1-\epsilon_1(a_c))^2}\left(2-\epsilon_1(a_c)+3\epsilon_3(a_c)\right)\,
.
\end{equation}
The spectral index $n_{\mathcal{S}}$ in terms of the parameter
$n_s$ is \cite{reviews1,Hwang:2005hb},
\begin{equation}\label{spectralindexprofinal}
n_{\mathcal{S}}=4-\sqrt{4n_s+1}\, ,
\end{equation}
while the spectral index of the tensor perturbations is given by,
\begin{equation}\label{spectralindexprofinal}
n_{T}=3-\sqrt{4n_t+1}\, ,
\end{equation}
so by using Eqs. (\ref{nsparameter}) and (\ref{ntparameter}) and
by expanding the square roots for $\epsilon_i(a_c)\ll 1$ we get
the final expressions of the spectral indices for $F(R)$ gravity,
\cite{reviews1,Hwang:2005hb},
\begin{equation}
\label{epsilonall} n_{\mathcal{S}}=
1-\frac{4\epsilon_1(a_c)-2\epsilon_3(a_c)+2\epsilon_4(a_c)}{1-\epsilon_1(a_c)},\quad
n_T=-\frac{\epsilon_1(a_c)-\epsilon_3(a_c)}{1-\epsilon_1(a_c)}\, ,
\end{equation}
while in a similar way, the tensor-to-scalar ratio at the horizon
crossing time instance is,
\begin{equation}\label{tensortoscalarratioathorizoncrossing}
r\simeq 48\frac{\epsilon_3^2(a_c)}{(1+\epsilon_3(a_c))^2}\, .
\end{equation}
The Raychaudhuri equation for $F(R)$ gravity yields,
\begin{equation}\label{approx1}
\epsilon_1=-\epsilon_3(1-\epsilon_4)\, ,
\end{equation}
hence at leading order we have $\epsilon_1(a_c)\simeq
-\epsilon_3(a_c)$, and hence, the spectral index can be
approximated as follows,
\begin{equation}
\label{spectralfinal} n_{\mathcal{S}}\simeq
1-6\epsilon_1(a_c)-2\epsilon_4(a_c)\, ,
\end{equation}
and the tensor-to-scalar ratio becomes approximately,
\begin{equation}
\label{tensorfinal} r\simeq 48\epsilon_1(a_c)^2\, .
\end{equation}

Now we shall discuss an important issue, related to the fact that
the condition $\dot{\epsilon}_1(a_c)\neq 0$ holds true in our
case, and we discuss and compare how the conditions
$\dot{\epsilon}_1(a_c)\neq 0$ (used in this letter) and
$\dot{\epsilon}_1(a_c)= 0$ (used in the literature) affect the
physics related to early Universe $F(R)$ gravity phenomenology.
This discussion will clarify the importance of the theoretical
treatment presented in this letter, and the importance of
employing Karamata's theorem, instead of using the condition
$\dot{\epsilon}_1(a_c)= 0$, used in the literature
\cite{reviews1,reviews2,reviews3,reviews4,Hwang:2005hb,Stewart:1993bc}.
As we now demonstrate in a transparent way, the inconsistency in
$F(R)$ gravity inflation is strongly related with the slow-roll
parameter $\epsilon_4(a_c)$. A direct calculation of the parameter
$\epsilon_4$ for a general $F(R)$ gravity yields,
\begin{equation}\label{epsilon4final}
\epsilon_4(a_c)\simeq -\frac{24
F_{RRR}(a_c)H_c^2}{F_{RR}(a_c)}\epsilon_1(a_c)-3\epsilon_1(a_c)+\frac{\dot{\epsilon}_1(a_c)}{H_c\epsilon_1(a_c)}\,
.
\end{equation}
But if the condition $\dot{\epsilon}_1(a_c)\neq 0$ holds true, the
term $\dot{\epsilon}_1(a_c)$ under the slow-roll assumptions,
would read,
\begin{equation}\label{epsilon1newfiles}
\dot{\epsilon}_1(a_c)=-\frac{\ddot{H}(a_c)H_c^2-2\dot{H}^2(a_c)H_c}{H^4(a_c)}=-\frac{\ddot{H}(a_c)}{H_c^2}+\frac{2\dot{H}^2(a_c)}{H^3(a_c)}\simeq
2H_c \epsilon_1^2(a_c)\, ,
\end{equation}
therefore the final approximate expression of $\epsilon_4(a_c)$
is,
\begin{equation}\label{finalapproxepsilon4}
\epsilon_4(a_c)\simeq -\frac{24
F_{RRR}(a_c)H_c^2}{F_{RR}(a_c)}\epsilon_1-\epsilon_1(a_c)\, .
\end{equation}
Let us note that in the case that the condition
$\dot{\epsilon}_1(a_c)=0$ is imposed, then this would lead many
well-known viable $F(R)$ gravity models, to become non-viable.
Indeed, the condition $\dot{\epsilon}_1(a_c)=0$ or equivalently
$\epsilon_1=0$ does not even apply for most of the well known
$F(R)$ gravities and in fact it applies only for power-law $F(R)$
gravity of the form $F(R)\sim R^n$, with $1<n<2$, but does not
apply for exponential models, or even the $R^2$ model. Let us
discuss for example the $R^2$ model \cite{Starobinsky:1980te},
which has $\epsilon_1(a_c)\sim \frac{1}{2N}$ at leading order in
the $e$-foldings number $N$, so if we assume the previous strong
condition $\dot{\epsilon}_1(a_c)=0$, and apply it in Eq.
(\ref{epsilon4final}) which holds true for a general $F(R)$
gravity, then the parameter $\epsilon_4(a_c)$ would be,
\begin{equation}\label{epsilon4finalr2}
\epsilon_4(a_c)\simeq -\frac{24
F_{RRR}(a_c)H_c^2}{F_{RR}(a_c)}\epsilon_1(a_c)-3\epsilon_1(a_c)+\frac{\dot{\epsilon}_1(a_c)}{H_c\epsilon_1(a_c)}=-3\epsilon_1(a_c)\,
,
\end{equation}
so the spectral index would be,
\begin{equation}
\label{epsilonallr2model} n_{\mathcal{S}}\simeq
1-6\epsilon_1(a_c)-2\epsilon_4(a_c)=1\, .
\end{equation}
This result would lead to a major inconsistency, since the $R^2$
model would result to a scale invariant power spectrum, and thus
would be excluded from the observational data, which is not true
since the $R^2$ model is perfectly consistent with the
observational data. However, by using the present theoretical
framework developed in this paper, if the condition
$\dot{\epsilon}_1\neq 0$ holds true, then the slow-roll index
$\epsilon_4$ can be calculated by using Eq.
(\ref{finalapproxepsilon4}), so it would be $\epsilon_4(a_c)\sim
-\epsilon_1(a_c)$ and therefore the spectral index would be
approximately equal to,
\begin{equation}
\label{epsilonallr2model2} n_{\mathcal{S}}\simeq
1-4\epsilon_1(a_c)\, ,
\end{equation}
and since for the $R^2$ model the slow-roll index
$\epsilon_1(a_c)\simeq \frac{1}{2N}$, the spectral index reads,
\begin{equation}
\label{epsilonallr2model3} n_{\mathcal{S}}\simeq 1-\frac{2}{N}\, ,
\end{equation}
which is the well-known result for the $R^2$ model
\cite{reviews1,reviews2,reviews3,reviews4}.

Thus with this short letter, we employed Karamata's theorem for
regularly varying functions, in order to rectify the inconsistency
in the derivation of the spectral indices of $F(R)$ gravity, from
the power spectrum, which in the previous literature relies
heavily on the condition $\dot{\epsilon_1}=0$ or equivalently
$\epsilon_1=$const. By using solely the slow-roll conditions
$\epsilon_i(\alpha_{c})\ll 1$ at horizon crossing, and also by
further assuming that the terms $\sim \dot{\epsilon}_i(a_c)$,
$\sim H_c^2\epsilon_i(a_c)\epsilon_j(a_c)$, $\sim
\ddot{\epsilon}_i(a_c)$ and $\sim \dot{\epsilon}_i^2(a_c)$ are
subdominant in comparison to the terms $H_c^2\epsilon_i(a_c)$ at
horizon crossing, we demonstrated that in view of Karamata's
theorem, the final expressions of the spectral index of primordial
scalar curvature perturbations $n_{\mathcal{S}}$, the tensor
spectral index $n_T$, and of the tensor-to-scalar ratio $r$
appearing in many reviews and textbooks in the literature, are
actually valid, without assuming that $\dot{\epsilon}_1(a_c)=0$,
thus rectifying the derivation of the resulting formulas and
refining the argument. We need to note that the method we used in
this letter for the derivation of the spectral indices is directly
applicable for any $f(R,\phi)$ theory and even for $f(R,\phi,K)$
theories with Einstein-Gauss-Bonnet corrections and tachyonic
corrections.

Before closing let us further clarify the important outcomes of
this  work. In the present existing literature related to $F(R)$
gravity inflationary phenomenology, the spectral index of the
primordial scalar curvature perturbations $n_{\mathcal{S}}$ and
the tensor to scalar ratio $r$, have the expressions given in Eqs.
(\ref{spectralfinal}) and (\ref{tensorfinal}) respectively, namely
$n_{\mathcal{S}}\simeq 1-6\epsilon_1(a_c)-2\epsilon_4(a_c)$, and
$r\simeq 48\epsilon_1^2$. These expressions are exactly the same
with the ones we obtained in this work, with the important
difference that in the present literature
\cite{reviews1,reviews2,reviews3,reviews4,Hwang:2005hb}, without
the use of Karamata's theorem, the expressions
(\ref{spectralfinal}) and (\ref{tensorfinal}) were obtained by
assuming that $\dot{\epsilon_1}=0$ or equivalently
$\epsilon_1=$const. As we showed however, the condition
$\epsilon_1=$const, also used in the context of minimally coupled
scalar theory \cite{Stewart:1993bc}, is not necessary and in fact
the same applies for $F(R)$ theory, by making use of the slow-roll
conditions imposed on the slow-roll indices, and the fact that
during inflation the slow-roll indices are slowly varying.

Finally, let us note that the refinement of the $F(R)$ gravity
inflation that we presented in this letter, does not give any new
information on how viable $F(R)$ gravities should be chosen.
However, we have strong indications that a universal relation
exists that governs all the viable $F(R)$ gravities, and this is
based on the condition $\dot{\epsilon}_1\neq 0$ for these
theories. We shall present the results soon \cite{odioiko}.

\end{document}